\documentclass{elsart}

\usepackage{graphicx}

\usepackage{amssymb}

\newcounter{bla}
\newenvironment{refnummer}{%
\list{[\arabic{bla}]}%
{\usecounter{bla}%
 \setlength{\itemindent}{0pt}%
 \setlength{\topsep}{0pt}%
 \setlength{\itemsep}{0pt}%
 \setlength{\labelsep}{2pt}%
 \setlength{\listparindent}{0pt}%
 \settowidth{\labelwidth}{[9]}%
 \setlength{\leftmargin}{\labelwidth}%
 \addtolength{\leftmargin}{\labelsep}%
 \setlength{\rightmargin}{0pt}}}
 {\endlist}

\begin{document}
\begin{frontmatter}

\title{New version of PLNoise: a package for exact numerical simulation of power-law noises}

\author{Edoardo Milotti\corauthref{cor1}}
\ead{milotti@ts.infn.it}
\corauth[cor1]{tel. +39 040 558 3388, fax +39 40 558 3350}
 
\address{Dipartimento di Fisica, Universit\`a di Trieste and I.N.F.N. -- Sezione di Trieste \\ Via Valerio, 2 -- I-34127 Trieste, Italy }




\begin{abstract}
In a recent paper I have introduced a package for the exact simulation of power-law noises and other colored noises (E. Milotti, Comput. Phys. Commun. {\bf 175} (2006) 212): in particular the algorithm generates $1/f^\alpha$ noises with $0 < \alpha \leq 2$. Here I extend the algorithm to generate $1/f^\alpha$ noises with $2 < \alpha \leq 4$ (black noises). The method is exact in the sense that it produces a sampled process with a theoretically guaranteed range-limited power-law spectrum for any arbitrary sequence of sampling intervals, i.e., the sampling times may be unevenly spaced.
\begin{flushleft}


\end{flushleft}


\end{abstract}

\end{frontmatter}


{\bf NEW VERSION PROGRAM SUMMARY}

\begin{small}
\noindent
{\em Program Title:} PLNoise \\                           
{\em Catalogue identifier:}                                   \\ \\
{\em Licensing provisions:}  none                                 \\  \\
{\em Programming language:} ANSI C. \\
\\
{\em Computer:} Any computer with an ANSI C compiler: the package has been tested with gcc version 3.2.3 on Red Hat Linux 3.2.3-52 and gcc version 4.0.0 and 4.0.1 on Apple Mac OS X-10.4.\\
 \\
{\em Operating system: } All operating systems capable of running an ANSI C compiler.  \\
 \\
{\em RAM:} the code of the test program is very compact (about 60 Kbytes), but the program works with list management and allocates memory dynamically; in a typical run with average list length $2 \cdot 10^4$, the RAM taken by the list is 200 Kbytes. \\                           
\\
{\em Classification:} 4.13 \\                              
\\
{\em External routines:} The package needs external routines to generate uniform and exponential deviates. The implementation described here uses the random number generation library {\small{\tt ranlib}} freely available from Netlib [1], but it has also been successfully tested with the random number routines in Numerical Recipes [2]. Notice that {\small{\tt ranlib}} requires a pair of routines from the linear algebra package {\small{\tt LINPACK}}, and that the distribution of {\small{\tt ranlib}} includes the C source of these routines, in case {\small{\tt LINPACK}} is not installed on the target machine.  \\
\\
{\em Catalogue identifier of previous version:}  ADXV\_v1\_0             \\
\\ 
{\em Journal reference of previous version:}  Comput. Phys. Commun. {\bf 175} (2006) 212  \\
\\ 
{\em Does the new version supersede the previous version?:}  Yes  \\
\\
{\em Nature of problem:} Exact generation of different types of colored noise.\\
   \\
{\em Solution method:} Random superposition of relaxation processes [3], possibly followed by an integration step to produce noise with spectral index $> 2$. \\
\\
 {\em Reasons for the new version:} Extension to $1/f^\alpha$ noises with spectral index $2<\alpha \le 4$: the new version generates both noises with spectral with spectral index $0<\alpha \le 2$ and with $2<\alpha \le 4$.\\
   \\
{\em Summary of revisions:} although the overall structure remains the same, one routine has been added and several changes have been made throughout the code to include the new integration step.  \\
   \\
{\em Unusual features:} The algorithm is theoretically guaranteed to be exact, and unlike all other existing generators it can generate samples with uneven spacing.\\
   \\
{\em Additional comments:} The program requires an initialization step; for some parameter sets this may become rather heavy.\\
   \\
{\em Running time:} running time varies widely with different input parameters, however in a test run like the one in section \ref{desc} in the long write-up, the generation routine took on average about 75 $\mu$s for each sample. \\
   \\
{\em References:}
\begin{refnummer}
\item B. W. Brown, J. Lovato, and K. Russell: {\small{\tt ranlib}}, available from Netlib \\
({\small{\tt http://www.netlib.org/random/index.html}}, select the C version {\small{\tt ranlib.c}}).
\item W. H. Press, S. A. Teulkolsky, W. T. Vetterling, and B. P. Flannery, {\it Numerical Recipes in C: The Art of Scientific Computing, 2nd ed.} pp. 274-290 (Cambridge Univ. Press., Cambridge, 1992).
\item E. Milotti, Phys. Rev. E {\bf 72}, 056701 (2005).
\end{refnummer}

\end{small}

\newpage


\hspace{1pc}
{\bf LONG WRITE-UP}

\section{Introduction}
\label{intro}

I have recently developed a power-law noise generator which uses a superposition of uncorrelated simple relaxation processes with a uniform distribution of relaxation rates to produce $1/f$ noise, and a power-law distribution of relaxation rates to obtain $1/f^\alpha$ noise \cite{milotti,milotti2}: the generator can be easily modified to produce complex superpositions of different power-law noises, like those observed in precise clocks \cite{vessot}. This generator also has a very unusual and unique feature: the superposition mechanism takes into account the correlation between different samples, and does it so {\em exactly}, so that its noise output can be sampled at arbitrary sampling times. This is important whenever sampling cannot be performed at evenly spaced times: uneven sampling is quite common in astronomical observations \cite{ado}, but also in other fields such as climatology \cite{hd}, it shows up whenever observations have unintentional gaps and missing data \cite{wilson}, and sometimes bunched sampling may even be a technical need, e.g. in hard-disk controllers \cite{wu}; moreover, uneven sampling has some special properties, because it is not band limited in the same sense as ordinary, even sampling \cite{beu}. The generator does not produce exactly Gaussian noise, but the closeness to Gaussianity is tunable and the noise distribution can approach a true Gaussian to any required degree by a proper parameter setting \cite{milotti}. 

Most of the observed power-law ($1/f^\alpha$) noises have spectral indexes  $0 < \alpha \leq 2$, with an apparent clustering around $\alpha = 1$. However noises with higher spectral indexes $2 < \alpha \leq 4$ also show up in several unrelated systems \cite{mandel,schroeder} like the water level of the Nile river, economics, orchid population size \cite{gillman} and local temperature fluctuations and affect precise timekeeping \cite{vessot} and our ability to predict environmental and animal population variables \cite{kim}. Noises with $\alpha > 2$ also appear in the energy level fluctuations of quantum systems \cite{rel,sal}. These noises correspond to generalizations of the standard one-dimensional random walk, and for this reason the corresponding process is often called {\it fractional Brownian motion} (fBm) (likewise Gaussian noises with $0 \leq \alpha \leq 2$ are also called {\it fractional Gaussian noises} (fGn)). Because of their extreme peaking behavior at low frequencies  these noises are also called ``black'' \cite{schroeder}, and they display marked persistence properties \cite{mandel} that may lead to the mistaken identification of underlying trends in experimental data \cite{rang}.

It is easy to see that there is no way to produce black noises from the superposition of simple relaxation processes, the spectral density of a train of random pulses with simple exponential response with decay rate $\lambda$ is proportional to $1/(\omega^2+\lambda^2)$ and its derivative in a log-log plot ranges from 0 to -2: thus no simple superposition can have a spectrum that decays faster than $1/f^2$ and the algorithm described in \cite{milotti,milotti2} is unable to simulate this very interesting class of power-law noises. This paper addresses the practical problem of including black noises in the the noise generator introduced in \cite{milotti2}, preserving all its good features.

\section{Generation of black noises}
\label{method}

It may be argued that the superposition argument could still be used if one takes pulses with non-exponential shapes, however in that case the theory exposed in \cite{milotti} must be revised and the analysis becomes considerably harder. These difficulties can be circumvented when we note that the relationship between the spectral density $S_x(\omega)$ of a random process $x$ and the spectral density $S_y(\omega)$ of its integral $y$ is $S_y(\omega)=S_x(\omega)/\omega^2$ and therefore it is possible to generate a black noise taking the integral of a power-law noise obtained from simple superposition. Here, as in \cite{milotti,milotti2}, I consider a noise signal $x(t)$ which is a linear superposition of many random pulses, i.e., pulses that are random in time and are associated to a memoryless process with a Poisson distribution, and such that their pulse response function is 
\begin{equation}
\label{prf}
h(t,\lambda) = 
\left\{
\begin{array}{cc}
\exp(-\lambda t)  & \mathrm{ if } \; t \ge 0\\
0  &  \mathrm{ if } \; t<0
\end{array}
\right.
\end{equation}
with a decay rate $\lambda$ which is a positive definite random variate with probability density $g_\lambda(\lambda)$ in the range $(\lambda_{min},\lambda_{max})$ so that
\begin{equation}
\label{sum}
x(t) = \sum_k A_k h(t-t_k, \lambda_k)
\end{equation}
where $t_k$ is the time at which the $k$-th pulse occurs, $A_k$ is its amplitude and $\lambda_k$ is the decay rate of its pulse response function. In principle the amplitude is also a random variate, but here I take a fixed amplitude $A_k = A$. If $n$ is the pulse rate, it can be shown \cite{milotti} that the average signal level is 
\begin{equation}
\label{mean2}
\langle x \rangle  =  n A  \left\langle \frac{1}{\lambda} \right\rangle ,
\end{equation}
the variance is 
\begin{equation}
\label{msq2}
\langle (\Delta x)^2 \rangle  =  \frac{n A^2}{2}  \left\langle \frac{1}{\lambda} \right\rangle ,
\end{equation}
and the spectral density is 
\begin{equation}
\label{psd2}
S(\omega) = \frac{n A^2}{2\pi} \int_{\lambda_{min}}^{\lambda_{max}} \frac{g_\lambda(\lambda)}{\omega^2 + \lambda^2} d\lambda ,
\end{equation}
where the decay rate $\lambda$ which is a positive definite random variate with probability density $g_\lambda(\lambda)$ in the range $(\lambda_{min},\lambda_{max})$. I define the normalized zero-mean process as follows
\begin{equation}
x_N(t) = \frac{x(t) - \langle x \rangle }{\sqrt{\langle (\Delta x)^2 \rangle}}=\frac{A\sum_k  h(t-t_k, \lambda_k) - \langle x \rangle }{\sqrt{\langle (\Delta x)^2 \rangle}} ,
\end{equation}
and the integral of $x_N(t)$ is
\begin{equation}
\label{ydef}
y_N(t)= \int_{-\infty}^{t}x_N(t')dt'  = \int_{-\infty}^{t}\frac{x(t') - \langle x \rangle }{\sqrt{\langle (\Delta x)^2 \rangle}}dt' \end{equation} .
Equation (\ref{ydef}) is a trivial analytical answer which is still useless for inclusion  in the generator and it must be suitably rearranged:  from equation (\ref{ydef}) we see that $\langle y_N(t) \rangle = 0$,  and that $y_N(t)$ is a kind of continuous one-dimensional random walk process
\begin{eqnarray}
y_N(t+\Delta t)& = & y_N(t) + \int_{t}^{t+\Delta t} x_N(t')dt'  \\
\label{ydef2}
& = & y_N(t) + \frac{1}{\sqrt{\langle (\Delta x)^2 \rangle}} \left( \int_{t}^{t+\Delta t} x(t')dt'  - \langle x \rangle \Delta t \right)
\end{eqnarray}
where $y_N(t)$ is the integrated process sampled at time $t$ and $y_N(t+\Delta t)$ is sampled at time $t+\Delta t$ (notice that $\Delta t$ {\em is not} a short integration time, but the arbitrary time interval between any two sampling times). From equation (\ref{sum}) we note that the integral in equation (\ref{ydef2}) can be rearranged as follows
\begin{eqnarray}
\nonumber
&& \int_{t}^{t+\Delta t} x(t')dt' = A\sum_k \int_{t}^{t+\Delta t}h(t'-t_k, \lambda_k)dt' \\
\label{p1}
&& =  A \sum_{t_k < t} \frac{\operatorname{e}^{-\lambda_k (t-t_k)}}{\lambda_k}\left[ 1- \operatorname{e}^{-\lambda_k \Delta t} \right] + A \sum_{t_k \in (t,t+\Delta t)} \frac{1}{\lambda_k}\left[ 1- \operatorname{e}^{-\lambda_k (t+\Delta t-t_k)} \right]
\end{eqnarray}
where the summations in (\ref{p1}) run over pulses that occured before $t$ and over those that occurred within the $(t,t+\Delta t)$ time interval. Finally from equations (\ref{ydef2}) and (\ref{p1}) we obtain a  formula for the integrated process $y_N$: 
\begin{eqnarray}
\nonumber
 y_N(t+\Delta t) &=& y_N(t) + \frac{1}{\sqrt{\langle (\Delta x)^2 \rangle}} \left\{ A \sum_{t_k < t} \frac{\operatorname{e}^{-\lambda_k (t-t_k)}}{\lambda_k}\left[ 1- \operatorname{e}^{-\lambda_k \Delta t} \right] \right. \\ 
\label{ok}
 && \left. + A \sum_{t_k \in (t,t+\Delta t)} \frac{1}{\lambda_k}\left[ 1- \operatorname{e}^{-\lambda_k (t+\Delta t-t_k)} \right] - \langle x \rangle \Delta t  \right\}
\end{eqnarray}
and this can be incorporated in the generator, because the summations can be evaluated using the events of the underlying Poisson process stored in a linked list as in \cite{milotti2}. Equation (\ref{ok}) is both a practical formula for the generation of the integrated process $y_N$ and a generalization of the equivalent updating formulas in \cite{gill} (those formulas hold only for the special case of the Ornstein-Uhlenbeck process and its integral, i.e. for $\alpha = 2$ and $\alpha =4$): the generation of the integrated process can be achieved producing and maintaining a sequence of Poisson distributed events as in \cite{milotti2} and using equation (\ref{ok}) instead of (\ref{sum}) to evaluate the noise process. Figures \ref{fig1} and figure \ref{fig2} show a pair of $x_N$, $y_N$ processes obtained with the methods described above, where the integrated process has a spectral index $\alpha = 3.5$. Figure \ref{fig3} shows the spectral density of the integrated process, and the numerical result is in excellent agreement with theory. 

Finally notice that when the input process $x(t)$ is Gaussian (and this can be achieved with  a proper choice  of generator parameters as explained in \cite{milotti}),  the difference $y_N(t+\Delta t) - y_N(t)$ is a Gaussian variate as well.

\section{Changes in the library}
\label{desc}

The changes in the library \cite{milotti2} are transparent for the user, and are scattered in several parts of  the code. 
As in the first version, the package contains two header files and two code files (plus the {\small{\tt ranlib}} files that are included for convenience, and are redistributed according to the standard Netlib rules). The first header file ({\small{\tt noise.h}}) contains the necessary {\small{\tt include}} statements and the structure definitions; the second header file ({\small{\tt noise\_prototypes.h}}) contains just the prototype definitions. 
In the present version of the generator, the structure used to share information between the generation routines (defined in the header file {\tt noise.h}) is
{\small
\begin{verbatim}
struct info
{

	double nt;				/* transition rate */
	double tau;				/* average transition time */

	double fillUpTime;		/* fill-up time estimate */
	double fillUpLength;	/* list length estimate */
	
	double average;			/* signal average and standard deviation */
	double sd;

	double meaninvlambda;	/* mean value of decay time */

	double lambdamin;		/* min and max decay rates */
	double lambdamax;
	double beta;			/* beta input by the user */
	double beta0;			/* actual value of beta used in the pulse distribution */
	double lmin;			/* aux. variables */
	double lmax;
	double dl;
	double binv;

};
\end{verbatim}
}
This header also contains the definition of $N_{decay}$: {\small{\tt \#define NDECAY 20.}}; with this definition the program is quite accurate (the average relative error after discarding the old events is $\approx 2\cdot 10^{-9}$), however the generation process can be made more accurate (and slower) with a larger value of {{\small{\tt NDECAY}}, or less accurate and faster with a smaller value. 

The first code file ({\small{\tt list\_routines.c}}) contains the code for the list routines: 
\begin{itemize}
\item {\small{\tt Append}}: this is a variant of the usual {\small{\tt Push}} routine;
\item {\small{\tt Process\_List}}: this routine processes the list to get rid of the old elements that can no longer influence the output signal;
\item {\small{\tt Print\_List}}: this routine prints the list;
\item {\small{\tt Response}}: this routine computes the noise signal for a spectral index between 0 and 2;
\item {\small{\tt IntegratedResponse}}: this routine computes the integrated noise signal and is called when the user asks for a spectral index between 2 and 4;
\item {\small{\tt Get\_List\_Length}}: this routine returns the length of the list.
\end{itemize}

these are internal routines, they are not meant to be called by the user, and are listed here for completeness; {\small{\tt IntegratedResponse}} is new in this version.

The second code file ({\small{\tt generator.c}}) contains the user-callable routines. These routines have the same names and calling sequences as the first version of the generator \cite{milotti2}, even though they have been modified in several places.

\section{Changes in the test program}

The user interface and the output file structure of the test program included in the package is the same as in the previous version \cite{milotti2}. The important difference is that now the allowed values of the spectral index $\alpha$ are in the range $0 \leq \alpha \leq 4$.

The output file begins with a header, which is a single line with the following values (separated by tabs):
\begin{enumerate}
\item {\small{\tt dt}} = sampling interval;
\item {\small{\tt nsamp}} = number of samples;
\item {\small{\tt tmax}} = time of last sample;
\item {\small{\tt noise\_info.nt}} = transition rate;
\item {\small{\tt noise\_info.tau}} = average time between transitions;
\item {\small{\tt noise\_info.lambdamin}} = $\lambda_{min}$;
\item {\small{\tt noise\_info.lambdamax}} = $\lambda_{max}$;
\item {\small{\tt noise\_info.beta}} = $\beta = \alpha -1$;
\item {\small{\tt noise\_info.meaninvlambda}} = $\langle 1/\lambda \rangle$;
\item {\small{\tt noise\_info.fillUpTime}} =  fill-up time estimate;
\item {\small{\tt noise\_info.fillUpLength}} =  list length estimate;
\item {\small{\tt noise\_info.average}} =  average output amplitude;
\item {\small{\tt noise\_info.sd}} =  standard deviation of noise signal;
\end{enumerate}
The rest of the file is a set of records, each with the following tab-separated values:
\begin{enumerate}
\item {\small{\tt kk}} = record number;
\item {\small{\tt t}} = actual time;
\item {\small{\tt tt}} = time of last transition event;
\item {\small{\tt listLength}} = list length;
\item {\small{\tt signal}} = signal (for $0 \leq \alpha \leq 2$) or integrated response (for $ 2 < \alpha \leq 4$);
\item {\small{\tt norm\_signal}} = normalized signal value;
\end{enumerate}

If $0 \leq \alpha \leq 2$ {\tt signal} is 
\begin{equation}
\label{response}
{\tt signal} = \sum_{t_k < t} \exp \left( \lambda_k (t - t_k) \right)
\end{equation}
and ${\tt norm\_signal} = x_N$, while if $ 2 < \alpha \leq 4$ {\tt signal} is the output of the {\tt IntegratedResponse} routine
\begin{equation}
\label{intresponse}
{\tt signal} = \sum_{t_k < t} \frac{\operatorname{e}^{-\lambda_k (t-t_k)}}{\lambda_k}\left[ 1- \operatorname{e}^{-\lambda_k \Delta t} \right] + \sum_{t_k \in (t,t+\Delta t)} \frac{1}{\lambda_k}\left[ 1- \operatorname{e}^{-\lambda_k (t+\Delta t-t_k)} \right]
\end{equation}
and ${\tt norm\_signal} = y_N$.

The code distribution also contains two {\it Mathematica} notebooks to analyze and display the program output; the notebook {\tt display.nb} is used for spectral index $0 \leq \alpha \leq 2$, while {\tt display2.nb} is used for spectral index $ 2 < \alpha \leq 4$.

The test program is used to generate the example discussed in the next section.

\section{Test run}
\label{test}

In the example shown below the spectral index is $\alpha = 3.5$ (i.e. the program generates $1/f^{3.5}$ noise, and $\beta = 2.5$), $\lambda_{min} = 0.0001$ and $\lambda_{max} = 1$ (the power-law region spans approximately 4 orders of magnitude), so that this corresponds very closely to the example given in \cite{milotti2}. And indeed, the program returns a signal which is the time integral of the sequence generated in the example given in \cite{milotti2}, and the underlying sequence is the same (the random number generator is the same as that in \cite{milotti2}, and uses the same seed for the random number sequence).

{\small
\begin{verbatim}
*** PLNoise ***

1. terminal input of control variables

Enter sampling time interval (dt): 1
Enter number of samples: 4194304
--> Total sampling time: 4.1943e+06
Enter transition rate: 0.1
--> Average transition time: 10
Enter lambda_min: 0.0001
Enter lambda_max (0 = single relax. rate, lambdamax = lambdamin): 1
Enter alpha (spectral index in 1/f^alpha): 3.5

2. initialization

Initialization time: 0.050000 seconds

Noise parameters: 

 -- Spectral shape: 
Min decay rate: 0.0001
Max decay rate: 1
Beta: 2.5 (spectral index is 1+beta = 3.5)

 -- Poisson process: 
Transition rate: 0.1
Average transition time: 10

 -- Algorithmic variables: 
Fill-up time: 200000
Fill-up length: 20000
Mean value of decay time (<1/lambda>): 100
Mean list length: 200
Signal average: 10
Signal variance: 5
Signal standard deviation: 2.23607
Signal skewness: 0.298142
Rule of thumb for Gaussianity: n<1/lambda> = 10 >= 10, noise is Gaussian

 -- Internal parameters: 
lmin: 0.01
lmax: 1
dl: 0.99
binv: 2

List length after initialization: 203

... initialized ... 
... starting now ...


3. main generation loop

|--------------------------------------------------|
 XXXXXXXXXXXXXXXXXXXXXXXXXXXXXXXXXXXXXXXXXXXXXXXXXX

Generation time: 313.780000 seconds

4. statistics

Statistics of generated sequence: 
Average: 10.0659
Variance: 4.91088
Standard deviation: 2.21605
Skewness: 0.318761

5. end
\end{verbatim}
}

The program has been compiled and run with the same compiler and on the same machine as the example in \cite{milotti2} (i.e., the compiler {\small{\tt gcc 4.0.1}} with optimization flag {\small{\tt -O3}} has been used, and the program has been run on an Apple Powerbook G4 - 1.5 GHz with the Mac OS X 10.4.6 UNIX flavor): a comparison between the generation times shows that in this case the integration step leads to a 75\% increase of generation time with respect to the unintegrated case.


\begin{figure}
\includegraphics[width=5in]{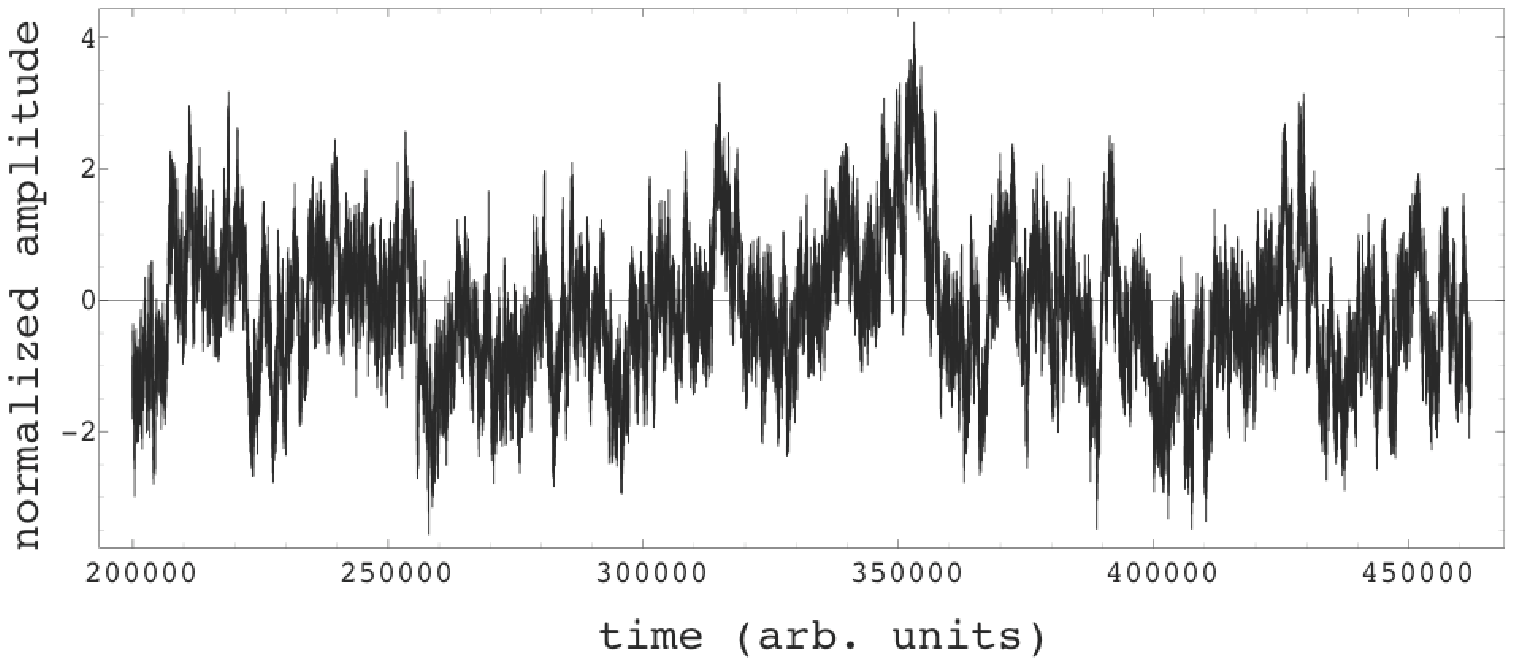}
\caption{\label{fig1} Normalized process $x_N(t)$ with power-law spectrum $1/f^{1.5}$ produced by the generator described in \cite{milotti,milotti2}. This record contains $2^{18} = 262144$ samples, and the generation parameters are $n=10$, $\lambda_{min} = 0.0001$, $\lambda_{max} = 1$, $\beta = 0.5$ (see \cite{milotti} for a detailed explanation of these parameters). Time does not start from 0 because the initial part of the noise record is used for initialization, and the relaxation rates $\lambda$ are given in arbitrary frequency units related to the arbitrary time units used in the figure. 
}
\end{figure}

\begin{figure}
\includegraphics[width=5in]{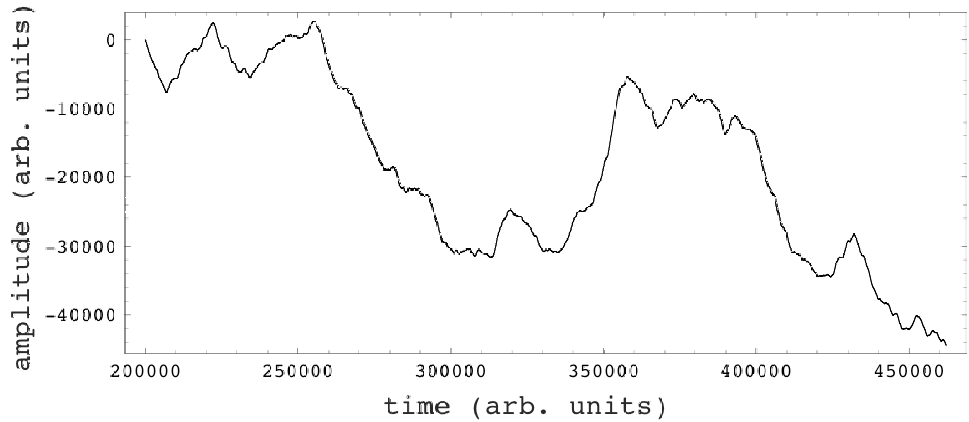}
\caption{\label{fig2} Plot of the process $y_N(t)$ obtained from the time integration of the process $x_N(t)$ shown in figure \ref{fig1}, as explained in the text. The apparent global linear trend is characteristic of black noise, and any such trend disappears  or is replaced by another different trend in longer records.
}
\end{figure}

\begin{figure}
\includegraphics[width=5in]{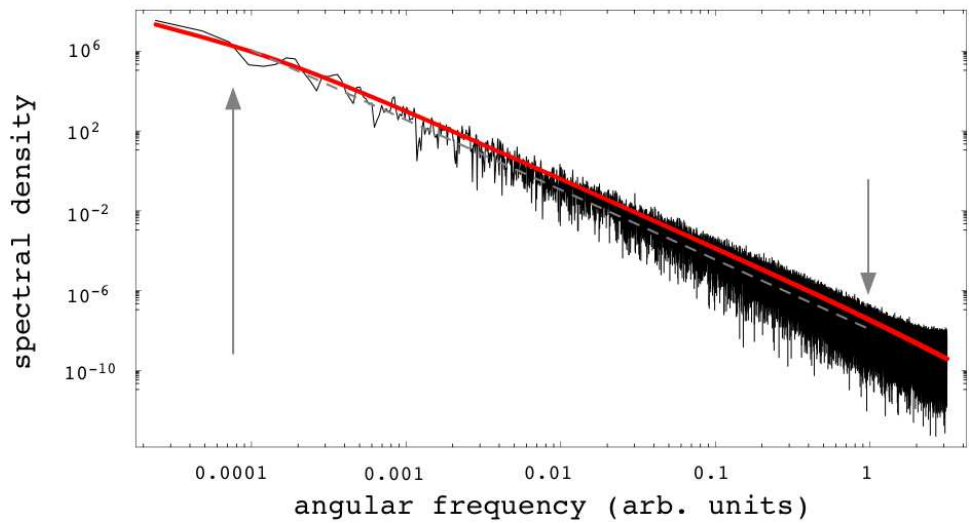}
\caption{\label{fig3} Spectral density of the process $y_N(t)$ shown in figure \ref{fig2}. The linear trend must be removed to avoid artifacts, either by detrending or by windowing: here I used a Hanning window. The arrows mark the positions of the minimum and maximum relaxation rates $\lambda_{min}$ and $\lambda_{max}$, the thick line is the expected average spectrum, corrected for the window incoherent gain, and the dashed line shows the expected slope of a $1/f^{3.5}$ spectral density. The slight upward bend at high frequency is due to uncorrected aliasing.}
\end{figure}

\end{document}